\documentclass[doublespacing]{elsart}

\usepackage{amssymb}

\begin{document}

\begin{frontmatter}


\title{Stabilization of Quantum Computer Calculation Basis by Qubit Encoding in Virtual Spin Representation}

\author{Alexander R. Kessel}
\ead{kessel@dionis.kfti.knc.ru}
\address{Kazan Physical-Technical Institute, Russian Academy of Science, \\ Sibirsky trakt 10/7, Kazan 420029, Russia}

\begin{abstract}
It is proposed to map the quantum information qubit not to individual spin 1/2 states, but to the collective spin states being eigenfunctions of the Hamiltonian including spin-spin interactions, which may be not small. Such an approach allows to introduce more stable calculation basis for quantum computer based on the solid state NMR systems.
\end{abstract}

\end{frontmatter}

\section{Introduction}

NMR is excellent testing ground for approbation of the different 
ideas of quantum information science. Due to the combination of 
the developed theory and the refined experimental technique it 
becomes to be possible to realize some algorithms on a few qubit 
quantum computer being implemented on standard NMR spectrometers. 
Up to day the quantum information science achievements based on 
liquid state NMR systems far exceed other experimental realization 
of quantum calculations.

The success of the liquid state quantum information science 
connected partially with the fact that it was found the clear 
and convenient physical object, nuclear spin $1/2$, for the 
representation of the quantum mechanical bit (qubit) conception. 
Two possible spin orientations in the dc magnetic field, which 
are described by the nuclear $z$-component eigenfunctions 
$|m\rangle(m=\pm1/2)$, are associated with two qubit states 
$|0\rangle$ and $|1\rangle$. The functions $|m\rangle$ are 
also eigenfunctions of the spin-system main Hamiltonian. 
Two qubit calculation basis arranged on the states of two spins is

\begin{eqnarray}
|00\rangle = |m_{1}=-1/2, m_{2}=-1/2\rangle, & |01\rangle=|m_{1}=-1/2, m_{2}=+1/2\rangle, \nonumber \\
|10\rangle = |m_{1}=+1/2, m_{2}=-1/2\rangle, & |11\rangle=|m_{1}=+1/2, m_{2}=+1/2\rangle. 
\label{eqn:1}
\end{eqnarray}

However the restrictions of liquid state NMR possibilities in 
quantum information science become evident yet now \cite{bib:1,bib:2}. 
The next step in development of information science based on 
NMR will be connected probably with solid state NMR. The 
essential feature of solid state NMR is the existence in 
the Hamiltonian of the spin-spin interactions (of exchange 
or dipole-dipole type), which is not averaged by the thermal 
motion and contains not only spin $Z$-components, but 
also $X$- and $Y$- components.

For the last reason the orientation of individual spin in 
the dc magnetic field in solid state media becomes bad integral 
of motion, that is the stationary states of spin system do not 
correspond to a definite value of the individual spin 
$Z$-component, and functions (\ref{eqn:1}) are unstable states. 
That is why the special efforts are required for damping 
the spin-spin interactions and for support the stability of 
calculation basis connected with individual spin orientation \cite{bib:2}. 
This can be reached by effecting every sort of spins in solid 
state by the pulse sequences developed in NMR, for example, 
by WAHUHA sequence \cite{bib:3} consisting of four resonance pulses.

Here we demonstrate that it is possible to achieve the 
increasing of the computational basis stability without 
damping the spin-spin coupling at the condition when spins 
$Z$-components do not commute with main Hamiltonian. 
The point is in the qubit mapping on a pair of the 
eigenstates of Hamiltonian, which include main 
Hamiltonian and spin-spin interaction, instead of 
eigenstates of the main Hamiltonian only, as it can 
be interpreted if functions (\ref{eqn:1}) are used for encoding.

In terms of spin orientations it means that virtual spin 
orientation is mapped to the qubit \cite{bib:4}. In this case the 
qubits turn out to be delocalized and direct correspondence 
between the qubit and individual spin is lost.

To make clear the idea of stabilization let us take in 
consideration a quantum system with the Hamiltonian 

\begin{displaymath}
\mathcal{H} = \mathcal{H}_{0} + V_{1} + V_{2} + V_{3} + \ldots ,
\end{displaymath}

where $\mathcal{H}_{0}$ is a main Hamiltonian of non interacting particles 
and $V_{1}\gg V_{2}\gg V_{3}\gg\ldots$ is a hierarchy of 
other interactions. Let the eigenfunctions of the main 
Hamiltonian are (\ref{eqn:1}).

In the case when qubits are encoded on the eigenstates 
$|m_{1}, m_{2}\rangle$ of main Hamiltonian, the qubit 
states remain stable only on the times $t_{1}\simeq h/V_{1}$. 
Then they vary under the influence of the $V_{1}$ interaction. 
The calculation basis stability will increase and will 
hold during the time interval $t_{2}\simeq h/V_{2}$, if the $V_{1}$ 
interaction is included in the main Hamiltonian and the 
qubits will be encoded on the eigenfunctions of the 
Hamiltonian $\mathcal{H}_{0}+V_{1}$. According to the 
definition of the interaction hierarchy it means: 
$t_{2}\gg t_{1}$. In other words the including of 
more and more weaker Hamiltonians in the qubit definition 
increases the basis stability. Such including needs the 
usage of qubit encoding in the virtual spin representation.

\section{A simple example: system of two interacting spins}

The complete set of gates, which is sufficient for forming 
the algorithm of an arbitrary complexity, consists of one 
qubit rotations and two qubit controlled negation gate 
CNOT \cite{bib:6}. That is why for an example it is enough to 
consider system of two nonequivalent coupled spins 
$I=1/2$ and $S=1/2$. The Hamiltonian of such a system is

\begin{eqnarray}
\mathbf{H}=\hbar\omega_{0}(\mathbf{I}_{z}+\mathbf{S}_{z})+\hbar\delta /2(\mathbf{I}_{z}-\mathbf{S}_{z})+\hbar J(\mathbf{I}\mathbf{S})+\hbar\mathbf{V}, \nonumber \\
\omega_{0}=(1/2)(\gamma_{I}+\gamma_{S})H_{0}, ~ \delta =-(\gamma_{I}+\gamma_{S})H_{0},
\label{eqn:2}
\end{eqnarray}

where $\gamma_{I}$ and $\gamma_{S}$ are the nuclear 
gyromagnetic ratio, $J$ is the exchange integral, $H_{0}$ 
is static magnetic field, $\hbar\mathbf{V}$ is the spin-lattice 
or dipole-dipole interaction. Let the inequalities 
$\omega_{0}>J>\mathbf{V}$ take place for the Hamiltonian parameters. 
The properties of this system are known for a long time \cite{bib:7}. 

This Hamiltonian eigenfunctions in case $\mathbf{V}=0$ are

\begin{eqnarray}
|\Psi_{1}\rangle = |++\rangle\equiv|m_{I}=+1/2,m_{S}=+1/2\rangle, & |\Psi_{2}\rangle = p|+-\rangle + q |-+\rangle, \nonumber \\
|\Psi_{3}\rangle = p|-+\rangle -q|+-\rangle, & |\Psi_{3}\rangle = |--\rangle,
\label{eqn:3}
\end{eqnarray}

where $p=\cos(\phi /2), q = \sin (\phi /2)$. The eigenvalues 

\begin{eqnarray}
E_{1}\equiv\hbar\varepsilon_{1}=\hbar\omega_{0}+(1/4)\hbar J, & E_{2}\equiv\hbar\varepsilon_{2}=-(1/4)\hbar J + (1/2)\hbar \theta, \nonumber \\
E_{4}\equiv\hbar\varepsilon_{4}=-\hbar\omega_{0}+(1/4)\hbar J, & E_{3}\equiv\hbar\varepsilon_{3}=-(1/4)\hbar J - (1/2)\hbar \theta, 
\label{eqn:4}
\end{eqnarray}

correspond to these eigenfunctions (\ref{eqn:3}), where 
$\theta^{2} = J^{2} + \delta^{2}$. In such a system there 
are four allowed resonance transitions on the frequencies 

\begin{eqnarray}
\varepsilon_{12}=\omega_{0}+(1/4)J-(1/2)\theta, & \varepsilon_{13}=\omega_{0}+(1/4)J+(1/2)\theta, \nonumber \\
\varepsilon_{24}=\omega_{0}-(1/4)J-(1/2)\theta, & \varepsilon_{34}=\omega_{0}-(1/4)J-(1/2)\theta,
\label{eqn:5}
\end{eqnarray}

having the relative intensities 

\begin{eqnarray}
A_{24} = A_{12} \propto |\langle \Psi_{1}|I_{x}+S_{x}|\Psi_{2}\rangle |^{2} = 1 + \sin \phi, \nonumber \\
A_{34} = A_{13} \propto |\langle \Psi_{1}|I_{x}+S_{x}|\Psi_{3}\rangle |^{2} = 1 - \sin \phi.
\label{eqn:6}
\end{eqnarray}

\section{Calculation basis in virtual spin representation}

Four-dimensional Hilbert space spanned on the functions (\ref{eqn:3}) 
can be considered as a direct product of two two-dimensional 
Hilbert spaces of virtual spins $I=1/2$ and $S=1/2$ \cite{bib:8}. It means 
that the Hamiltonian (\ref{eqn:2}) eigenfunctions (\ref{eqn:3}) are taken as the 
calculation basis, that is

\begin{eqnarray}
|00\rangle = |\Psi_{1}\rangle, & |01\rangle = |\Psi_{2}\rangle, \nonumber \\
|10\rangle = |\Psi_{3}\rangle, & |11\rangle = |\Psi_{4}\rangle,
\label{eqn:7}
\end{eqnarray}

In such notations the separate virtual spin corresponds to the separate qubit, that is 

\begin{eqnarray*}
|0\rangle = |m_{Q}= -1/2\rangle, |1\rangle = |m_{Q}=+1/2\rangle, \\
|0\rangle = |m_{R}= -1/2\rangle, |1\rangle = |m_{R}=+1/2\rangle, 
\end{eqnarray*}

and coupled states of two qubits form the calculation basis

\begin{eqnarray}
|00\rangle = |m_{Q}=-1/2, m_{R}=-1/2\rangle, & |01\rangle = |m_{Q}=-1/2, m_{R}=+1/2\rangle, \nonumber \\
|10\rangle = |m_{Q}=+1/2, m_{R}=-1/2\rangle, & |11\rangle = |m_{Q}=+1/2, m_{R}=+1/2\rangle, 
\label{eqn:8}
\end{eqnarray}

where $m_{Q}$ and $m_{R}$ are the "eigenvalues" of the virtual 
spins $Q=1/2$ and $R=1/2$. The functions (\ref{eqn:8}) are connected to 
the Hamiltonian (\ref{eqn:2}) eigenstates (\ref{eqn:3}) by relations (\ref{eqn:7}).

Now the resonance transition between two states of the real physical 
system attributed to a single qubit can be interpreted as a virtual 
spin reorientation. For example, the transition 
$\langle\Psi_{1}|\leftrightarrow|\Psi_{2}\rangle$ can be interpreted 
in virtual spin representation as a virtual spin $R$ rotation and 
so on. It can be shown (see the Table \ref{tab:1}) using the results of 
papers \cite{bib:4,bib:5} that universal gates being established in the two 
qubit system under consideration may be realized by suitable 
pulses of resonance radio frequency field. 

\begin{table}
\caption{Logic operations and implementation pulses}
\begin{tabular}{|l|l|}
\hline
Logic operation & Excitation transitions \\
\hline
Virtual spin $Q$ rotation & $\langle\Psi_{1}|\leftrightarrow|\Psi_{2}\rangle$ and $\langle\Psi_{3}|\leftrightarrow|\Psi_{3}\rangle$ \\
Virtual spin $R$ rotation & $\langle\Psi_{1}|\leftrightarrow|\Psi_{3}\rangle$ and $\langle\Psi_{2}|\leftrightarrow|\Psi_{4}\rangle$\\
Controlled negation of spin $Q$  & $\pi$-pulse on the $\langle\Psi_{3}|\leftrightarrow|\Psi_{4}\rangle$ transition \\
CNOT$_{R\rightarrow Q}$ & \\
Controlled negation of spin $R$ & $\pi$-pulse on the $\langle\Psi_{2}|\leftrightarrow|\Psi_{4}\rangle$ transition \\
CNOT$_{Q\rightarrow R}$ & \\
\hline
\end{tabular}
\label{tab:1}
\end{table}

These quantum gates in the two qubit quantum system can be 
realized by one double frequency pulse for qubit rotation 
and one single frequency pulse for controlled negation \cite{bib:5}.

\section{Conclusion}
As it was shown above the information encoding in the virtual 
spin representation permits to create rather simple the 
universal set of quantum gates in system of two interacting 
real spins located in the solid state. The preference of 
this type encoding lies in the possibility to use systems 
of two spins with large spin coupling and in the absence of 
necessity to subject the solid state to continuous pulse 
influence for supporting the calculation basis stability. 
The basis stability in system under consideration will 
hold during time interval $t_{1}\simeq 1/J$, if qubit encoding is 
performed on the eigenstates (\ref{eqn:1}) of Zeeman Hamiltonian. 
The stability spreads to a larger interval $t_{2}\simeq 1/J$, 
if the encoding is established on the eigenstates (\ref{eqn:3}) 
of the Hamiltonian (\ref{eqn:2}) being considered without spin-lattice 
and dipole-dipole interaction. Besides the operation time of 
such a gate is in the whole experimentalists control and can 
be made sufficiently short, whereas the operation time when 
coding is fulfilled on the real spins is determined by the 
exchange interaction value (by substance properties) \cite{bib:6} and, 
generally speaking, may be long.

The proposed approach is applicable to qubit encoding 
in the quantum information medium of an arbitrary nature, 
if there are suitable selection rules for external excitation, 
which is necessary for gate arrangement. In particular it may 
be the cluster of strong interacting particles or 
liquid state NMR systems in the case, where the exchange 
interaction is not averaged to the spin $z$-components. 
An example of virtual qubits encoding on the optical 
states of an individual atom is given in \cite{bib:8}.

\begin{ack}
This work was supported by REC -007 and fond NIOKR RT 06-6.1-158.
\end{ack}

\end{document}